\documentstyle[bezier,11pt,fleqn]{article}
\title{Note on the 8$_{18}$-Knot}
\begin{document}           
\noindent
{\Large \hspace*{52mm}{\bf Note on the 8$_{18}$-Knot}}

\vspace{4mm}
\noindent
{\large \hspace*{60mm}{A. Kwang-Hua Chu}}  \newline

\noindent
\hspace*{35mm}{\small P.O. Box 39, Tou-Di-Ban, Road XiHong, Urumqi 830000, PR China}
\begin{abstract}
We define some physical variables associated with the traversing sequences
of electrons along the orbit which is a 2D projection of 8$_{18}$-knot.
The configuration is regular but the resulting contributions, which are related
to the physical variable, of those combinations from all the possible states
to the fixed spatial sites show certain irregular behavior near the
over- or under-crossing points of this knot. The possible explanation
for this kind of direct geometric consequences is made to linked to the
physical insight.

\vspace{3mm}
\end{abstract}
\doublerulesep=4mm        
\baselineskip=6.5mm
\oddsidemargin+1mm         
\bibliographystyle{plain}
\section{Introduction}
Knots are classified according to the minimum number of intersections on
their projections \cite{Burde:knot}. Their applications to statistical
mechanics or topology of polymers and DNA dated back to early 70s
\cite{StM:Polymer}.
Knots are usually categorized in terms of topological properties that
are invariant under changes in a knot's spatial configuration [3,4].
For the special interests of our study here, we will investigate certain
property of 8$_{18}$-knot [5] which has configurationally similar but
topologically distinct relationship with some orbitals of electrons
\cite{Waldram:HTS}.
This knot has the polynomial invariant $\Delta(x)$
$=1-5x+10x^2-13x^3+10x^4-5x^5+x^6$ \cite{Alexander:1928} and zero writhing
number.  \newline
Recently studies of random knots and the knot probability in lattice
polygons have aroused many problems in the discrete form of knots or
lattice knots \cite{Nechaev:Random}. We shall give some preliminary
results about the discrete energy/entropy of 8$_{18}$-knot in this paper.
Our main focus will lie in the principles of indistinguishability and the
micro-reversibility related to the states of electrons with certain spin
after the electrons had traversed along this knot starting from any
preassigned original sites in its $xy$-projection within one cycle.
\section{General Approach}
Our idea is to consider the 8$_{18}$-knot as the traversing orbitals of
electrons which have certain spin. In the modeling of superconducting, e.g.
\cite{Scalapino:pairing}, the two pairing electrons avoid the Coulomb
interaction by arranging themselves in space in a $d_{x^2-y^2}$ orbit
which is similar to the 2D projection of 8$_{18}$-knot. We start from
the finite sequences of temporary electron-site which here are the crossings
and the vertices of its projection or 2D $xy$-configuration for simplicity.
These locations or moves of electron-site could be well defined in certain lattice.
There are actually 12 characteristic {\it points} as shown in Fig. (i). This
regular configuration is isotropic to any rotations in the $xy$-plane for the
consideration of 4 {\it branches} of the 8$_{18}$-knot. Starting from
any one of these 12 points and then traversing along this knot for one cycle,
the phase changes in 2D $xy$-projection are 6 $\pi$ with respect to the origin
o.  The center of an atom or a nucleus is located at the origin o.\newline
For simplicity, we treat those electrons moving within one cycle as a
system. The energy and entropy are assumed to be dependent on the size
of this system (as well as each element of this system, e.g., electron
with certain spin because of the spin-spin coupling, or when the
electrons are replaced by, vortices in the xy model \cite{K:T}), such as
the traversed area of this system in 2D case here. We can also imagine
the energy/entropy increment as the phase difference between the
temporary and original starting sites (we arbitrarily set) of the electrons.
\newline

\vspace{3mm}
\setlength{\unitlength}{1.00mm}   
\begin{picture}(120,44)(0,-6)
\thicklines
\put(51,30){\oval(14.2,15)[br]}
\put(57.8,29.2){\oval(15,14)[bl]}
{\bezier{40}(58,30)(58,38.6)(53.5,38.6)}
{\bezier{40}(50,31.5)(50,38.6)(54.5,38.6)}
\put(50,23.3){\oval(15.4,13.2)[tr]}
\put(56.8,22.5){\oval(13.2,14.6)[tl]}
\put(58,22.3){\circle*{1}}
\put(50,22.5){\circle*{1}}
\put(54.3,13.9){\circle*{1}}
\put(66.6,26){\circle*{1}}
\put(41.4,26){\circle*{1}}
\put(54,23){\circle*{1}}
\put(54,29.5){\circle*{1}}
\put(57,26){\circle*{1}}
\put(51,26){\circle*{1}}
\put(50,30){\circle*{1}}
\put(54,38.6){\circle*{1}}
\put(58,30){\circle*{1}}
{\bezier{40}(57.7,21)(58,13.9)(53.9,13.9)}
{\bezier{40}(50,22.5)(50,13.9)(54.1,13.9)}
{\bezier{40}(58,22.5)(66.6,22.5)(66.6,26)}
{\bezier{40}(59,30)(66.6,30)(66.6,26)}
{\bezier{40}(48.5,22.5)(41.4,22.5)(41.4,26)}
{\bezier{40}(50,30)(41.4,30)(41.4,26)}
\thinlines
\put(54,25.7){\makebox(0,0)[b]{\small o}}
\put(62,26){\makebox(0,0)[b]{\tiny x}}
\put(46,26){\makebox(0,0)[b]{\tiny x}}
\put(54,34){\makebox(0,0)[b]{\tiny x}}
\put(54,18){\makebox(0,0)[b]{\tiny x}}
{\bezier{30}(66.6,26)(66.6,38.6)(54,38.6)}
{\bezier{30}(66.6,26)(66.6,13.9)(54,13.9)}
{\bezier{30}(41.4,26)(41.4,38.6)(54,38.6)}
{\bezier{30}(41.4,26)(41.4,13.9)(54,13.9)}
\put(10,2){\makebox(0,0)[bl]{\small Fig. (i)\hspace*{1mm} Schematic
diagram of a {\it loosely} hard $8_{18}$-knot. Those points of solid
circle could be}}
\put(10,-2){\makebox(0,0)[bl]{\small \hspace*{12mm} prescribed as {\it
vertices} or {\it lattice sites}.}}
\end{picture}

There are 10 different states or cases for the 2D configuration ($xy$-projection of
the 8$_{18}$-knot here) in consideration because it is so regular and
isotropic as we have assumed. These states come from the starting sites
being at the center of each {\it branch} (e.g. I,J,K,L in Fig. 1), the outer
{\it shoulder} (e.g. E,F,G,H in Fig. 1), and the inner {\it shoulder} (e.g.
A,B,C,D in Fig. 1), respectively. As the object (i.e. electron here)
traverses, the area (or entropy) increases. We thus associate the moving
object (in a sequence) with the increasing area (or entropy) corresponding to
their orbital position (the same as the traversing time-sequence)
relative to the original starting position in our 2D configuration. For
convenience, we set the increment between any two of the 12 characteristic
points (in our observation) to be 1 (unit). The original starting value is 1.
But, those {\it shoulder} points, being the overcrossing or undercrossing with
respect to the same position in the $xy$-projection, must be indentified and
given the corresponding traversed values. \newline
Now, once the object is starting from the centers of the {\it branch}, it
has two choices (either clockwise or counterclockwise) to start traversing
because of the absent crossings. As the initial site moves to the outer and
inner {\it shoulders}, there are 4 choices for different traversing.
\section{Results and Discussions}
The results of the main different combinations are arranged and shown in Fig. 2
(a,b,c,d,e). Those values shown at the left-hand-side of "/" or upper
side of "--" come from the overcrossing situations whereas those shown at the
right-hand-side of "/" or lower side of "--" come from the undercrossing. We
neglect those mirrored cases ofthe 2D $xy$-projection of the 8$_{18}$-knot in
this presentation. We only plot those clockwise traversing cases which
are half of the 10 states within one revolution cycle in our
consideration. Thus there are missing 5 variants related to Fig. 2 (a,b,c,d,e)
corresponding to the counterclockwise traversing cases starting from
locations K, F, A, respectively. We can easily figure out these states
which are summarized in Table 1.\newline
The starting sites K-J-L-I/F-E-H-G/A-D-C-B, however, are indistinguishable
respectively because any rotations without deformations cannot change the
regularity of 2D $xy$-configuration we impose for this knot and besides,
there are no fixed labels for the sites. We only temporarily mark those
electron-sites with the 2D spatial projection-positions in the
$xy$-plane in order to observe the electron orbit conveniently and
instantly. The time sequences are artifically created by us and might not be the
real situations which, however, could be approximately emulated by
random processing subjected to certain physical rules if the sampling
numbers are large enough. That is to say, to start from position K makes
no differences with starting from position J or I or L. Similar
conclusions can also be made for those startings from the inner and outer
{\it shoulders}. \newline
After long-time averaging for all the possible states or traversing
sequences, the equilibrium contributions (entropy or interaction energy)
from the electron-motion to those 2D spatial sites with over- or under-crossing
should be almost the same if we accept that for the same group (center
of branch/outer shoulder/inner shoulder) since they are equally distant
from the origin o (there is the nucleus) the interactions should be nearly of
the same values.
But, from our results, as shown in Fig. 2, we can observe some mismatchs
by simply checking the addition of those values at the shoulders. We
have no idea about these kinds of {\it defects} related to the regular
8$_{18}$-knot.
These could be compensated only when the irregular hopping of the electron
happens around those {\it shoulders} of 8$_{18}$-knot during the traversing. We
make this statement by the direct geometric combination-results and
would like to wait for the explanations from the rigorous theories of physics
or well-measured experiments presently or in the future.

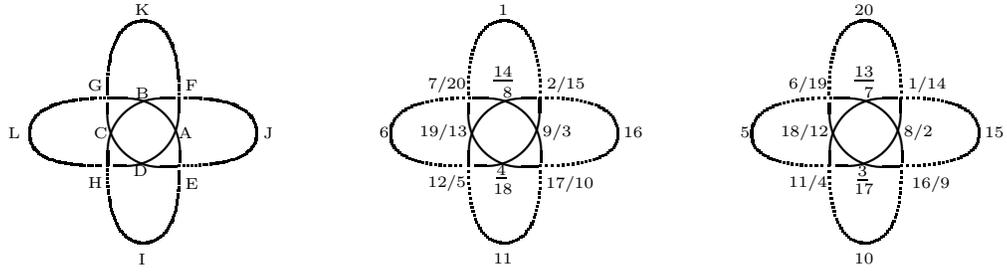
\end{document}